\documentclass[12pt]{article}

\usepackage{amsmath,graphicx}
\usepackage{amsfonts}
\usepackage{amssymb}
\usepackage{cite}
\usepackage{hyperref}

\def\hybrid{\topmargin -20pt
\oddsidemargin 15pt
\headheight 0pt
\headsep 0pt
\textwidth 6.25in       
\textheight 9in       
\marginparwidth .875in
\parskip 5pt plus 1pt
\jot = 1.5ex}

\hybrid
\numberwithin{equation}{section}
\numberwithin{table}{section}

\renewcommand{\Re}{\operatorname{Re}}
\renewcommand{\Im}{\operatorname{Im}}

\newcommand\e{\mathrm{e}}
\newcommand\iu{\operatorname{i}}
\newcommand\diff{\mathrm{d}}
\newcommand\su[1]{{\operatorname{SU}(#1)}}
\newcommand\so[1]{{\operatorname{SO}(#1)}}
\newcommand\M{{Y}}  
\newcommand\U{{U}}  
\newcommand\vol{{\operatorname{vol}}} 
\newcommand\tor{{t}}
\newcommand\torr{{t}}
\newcommand\Tor{{T}}

\def\Ei{{\mathcal{E}^i}}

\def\Eone{{\mathcal{E}^1}}
\def\Etwo{{\mathcal{E}^2}}
\def\Dc{{\mathcal{D}}}

\begin{document}

\begin{titlepage}

\begin{center}

\rightline{\small ZMP-HH/11-4}
\rightline{\small IPhT-t11/085}

\vskip 1cm

{\Large \bf The $\mathcal{N}=4$ effective action of type IIA supergravity compactified on $\su2$-structure manifolds }

\vskip 0.7cm

{\bf Thomas Danckaert$^{a}$, Jan Louis$^{a,b}$, Danny Mart\'inez-Pedrera$^{c}$,}\\
{\bf Bastiaan Spanjaard and Hagen Triendl$^{d}$}\\

\vskip 0.7cm

{}$^{a}${\em II. Institut f{\"u}r Theoretische Physik, 
Universit{\"a}t Hamburg,\\
Luruper Chaussee 149, 
D-22761 Hamburg, Germany}\\

\vskip 0.4cm

{}$^{b}${\em Zentrum f\"ur Mathematische Physik, Universit\"at Hamburg,\\
Bundesstrasse 55, D-20146 Hamburg, Germany}
\vskip 0.4cm

{}$^{c}${\em
Center for Quantum Engineering and Spacetime Research (QUEST) \&\\
Institut f\"ur Theoretische Physik, Leibniz Universit\"at Hannover\\
Appelstrasse 2, 30167 Hannover, Germany}\\

\vskip 0.4cm

{}$^{d}${\em Institut de Physique Th\'eorique, CEA Saclay,\\
Orme de Merisiers, F-91191 Gif-sur-Yvette, France}

\vskip 0.6cm

{\tt thomas.danckaert@desy.de, jan.louis@desy.de, danny.martinez@itp.uni-hannover.de, bastiaanspanjaard@googlemail.com, hagen.triendl@cea.fr} \\

\end{center}

\vskip 1cm

\begin{center} {\bf ABSTRACT } \end{center}

We study compactifications of type IIA supergravity on six-dimensional
manifolds with $\su2$ structure and compute the low-energy effective
action in terms of the non-trivial intrinsic torsion.
The consistency with gauged $\mathcal{N}=4$ supergravity is
established and the gauge group is determined. Depending on the structure of the intrinsic torsion, antisymmetric tensor fields can become massive.
\vfill

April 2011

\end{titlepage}


\tableofcontents

\section{Introduction}\label{section:Intro}

Compactification of ten-dimensional supergravities on generalized
manifolds with $G$-structure has been studied for some
time.\footnote{For reviews on this subject see, for example,
  \cite{Grana:2005jc,
Wecht:2007wu,Samtleben:2008pe} and references therein.} These manifolds are characterized by a reduced structure group $G$ which, when appropriately chosen, preserves part of the original ten-dimensional supersymmetry~\cite{Gauntlett:2002sc,Gauntlett:2003cy}. Furthermore, they generically have a non-trivial torsion which physically corresponds to gauge charges or mass parameters for some anti-symmetric tensor gauge potentials. Therefore, the low-energy effective action is a gauged or massive supergravity with a scalar potential which (partially) lifts the vacuum degeneracy present in conventional Calabi-Yau compactifications. The critical points of this scalar potential can further spontaneously break (some of) the left-over supercharges. As a consequence of this, such backgrounds are of interest both from a particle physics and a cosmological perspective.

Most studies so far concentrated on six-dimensional manifolds with
$\su{3}$ or more generally $\su{3}\times\su{3}$ structure. Compactifying the ten-dimensional heterotic/type~I supergravity on such manifolds leads to an $\mathcal{N}=1$ effective theory in four dimensions~\cite{Cardoso:2002hd,Becker:2003yv,Micu:2004tz,Gurrieri:2004dt,Benmachiche:2008ma}, while compactifying type II supergravity results in an $\mathcal{N}=2$ theory~\cite{Gurrieri:2002wz,Grana:2005ny,Grana:2006hr,Cassani:2007pq,Cassani:2008rb,Grana:2009im}. By employing an appropriate orientifold projection \cite{Benmachiche:2006df,Koerber:2007xk} or by means of spontaneous supersymmetry breaking \cite{Louis:2009xd,Louis:2010ui}, this $\mathcal{N}=2$ can be further broken to $\mathcal{N}=1$ (or $\mathcal{N}=0$).

A similar study for six-dimensional manifolds with $\su{2}$ or
$\su{2}\times \su{2}$ structure which generalize Calabi-Yau
compactifications on
$K3\times T^2$ has not been completed yet. In
Refs.~\cite{Gauntlett:2003cy,Bovy:2005qq,Triendl:2009ap}, geometrical
properties of such manifolds were studied and the scalar field space
was determined. Furthermore, it was shown in
Ref.~\cite{Triendl:2009ap} that manifolds with $\su{2}\times \su{2}$
structure cannot exist and therefore we only discuss the case of a
single $\su{2}$ in this paper. In Ref.~\cite{Louis:2009dq}, the
heterotic string was then compactified on manifolds with $\su{2}$
structure and the $\mathcal{N}=2$ low-energy effective action was
derived. In \cite{Danckaert:2009hr}, type IIA compactifications on
$\su{2}$ orientifolds were studied and again the corresponding
$\mathcal{N}=2$ effective action was determined. Finally in
Refs.~\cite{ReidEdwards:2008rd,Spanjaard:2008zz}, preliminary studies
of the $\mathcal{N}=4$ effective action for type IIA compactification
on manifolds with $\su{2}$ structure were conducted.\footnote{The
  effective action for IIA compactified on $K3\times T^2$ has been
  given in \cite{Duff:1995wd,Duff:1995sm}. $\mathcal{N}=4$ flux compactifications have been discussed for example in \cite{Aldazabal:2008zza,Dall'Agata:2009gv,Dibitetto:2010rg}.}

The purpose of this paper is to continue these studies and in particular determine the bosonic $\mathcal{N}=4$ effective action of the corresponding gauged supergravity. One of the technical difficulties arises from the fact that frequently in these compactifications magnetically charged multiplets and/or massive tensors appear in the low-energy spectrum. Fortunately, the most general $\mathcal{N}=4$ supergravity covering such cases has been determined in Ref.~\cite{Schon:2006kz} using the embedding tensor formalism of Ref.~\cite{deWit:2005ub}. We therefore rewrite the action obtained from a Kaluza-Klein (KK) reduction in a form which is consistent with the results of \cite{Schon:2006kz}. As we will see, this amounts to a number of field redefinitions and duality transformations in order to choose an appropriate symplectic frame.

The organization of this paper is as follows: In Section \ref{sec:SU2} we briefly review the relevant geometrical aspects of
$\su{2}$--structure manifolds and set the stage for carrying out the
compactification. Section~\ref{subsec:NS} deals with the reduction of
the NS-sector, which in fact coincides with the heterotic analysis
carried out in \cite{Louis:2009dq} and therefore we basically recall
their results. In Section~\ref{subsec:RR} we compactify the RR-sector
and give the effective action in  the KK-basis. In Section
\ref{sec:ConsistencyN=4} we perform the appropriate field
redefinitions and duality transformations in order to compare the
action with the results of Ref.~\cite{Schon:2006kz}. This allows us to
determine the components of the embedding tensor parametrizing the
$\mathcal{N}=4$ gauged supergravity action in terms of the intrinsic
torsion. From the embedding tensor we then can easily compute the gauge group in Section~\ref{Killing}.
Section~\ref{section:Conclude} contains our conclusions and some of the technical material is supplied  in the Appendices \ref{sec:dualizations-appendix} and \ref{sec:so6-n-coset}.

\section{SU(2) structures in six-manifolds} \label{sec:SU2}

\subsection{General setting} \label{sec:setting}

In this paper, we study type IIA space-time backgrounds of the form
\begin{equation}\label{background}
 M_{1,3} \times \M \ ,
\end{equation}
where $M_{1,3}$ denotes a four-dimensional Minkowski space-time and $\M$ a six-dimensional compact manifold.\footnote{Note that we do not consider warped compactifications in this work. For discussions of a non-trivial warp factor, see for instance~\cite{Koerber:2007xk,Martucci:2009sf}.} Furthermore, we focus on manifolds which preserve sixteen supercharges or in other words~$\mathcal{N}=4$ supersymmetry in four space-time dimensions. This implies that $\M$ admits two globally-defined nowhere-vanishing spinors $\eta^i$, $i=1,2$, that are linearly independent at each point of~$\M$. The necessity for this requirement can be most easily seen by considering the two ten-dimensional supersymmetry generators $\epsilon^1,\epsilon^2$, which are Majorana-Weyl and thus reside in the representation~$\mathbf{16}$ of the Lorentz group~$\so{1,9}$. For backgrounds of the form~\eqref{background}, the Lorentz group is reduced to~$\so{1,3} \times \so6$ and the spinor representation decomposes as
\begin{equation}
\mathbf{16} \to (\mathbf{2},\mathbf{4}) \oplus (\mathbf{\bar2},\mathbf{\bar4}) \ ,
\end{equation}
where $\mathbf{2}$ and $\mathbf{4}$ denote respectively four- and six-dimensional Weyl-spinor representations, while ${\mathbf{\bar2}}$ and ${\mathbf{\bar4}}$ are the corresponding conjugates. In terms of spinors we thus have
\begin{equation}
\begin{aligned}
\epsilon^1 & = \sum_{i=1}^2 (\xi^{1}_{i+} \otimes \eta^i_+ + \xi^{1}_{i-} \otimes \eta^i_-) \ , \\
\epsilon^2 & = \sum_{i=1}^2 (\xi^{2}_{i+} \otimes \eta^i_- + \xi^{2}_{i-} \otimes \eta^i_+) \ ,
\end{aligned}
\end{equation}
where the $\xi^{1,2}_i$ are the four $\mathcal{N}=4$ supersymmetry generators of $M_{1,3}$ and the subscript $\pm$ indicates both the four- and six-dimensional chiralities.

The existence of two nowhere-vanishing spinors $\eta^i$ forces the structure group of $\M$ to be $\su2$. This can be seen as follows. Recall that the spinor representation for a generic six-dimensional manifold is the fundamental representation $\mathbf{4}$ of $\su4 \simeq \so6$. The existence of two singlets implies the decomposition
\begin{equation}
\mathbf{4} \to \mathbf{2} \oplus \mathbf{1} \oplus \mathbf{1} \ ,
\end{equation}
which in turn leads to the fact that the structure group of the manifold is reduced to the subgroup acting on this $\mathbf{2}$, namely~$\su2$.


\subsection{Algebraic structure} \label{sec:algebraic}

Let us now briefly review the algebraic properties of $\su2$-structure manifolds. For a more detailed discussion, see~\cite{Triendl:2009ap}.

Instead of using the spinors $\eta^i$, we can parametrize the~$\su2$~structure on a six-dimensional manifold by means of a complex one-form~$K$, a real two-form~$J$ and a complex two-form~$\Omega$ \cite{Gauntlett:2003cy,Bovy:2005qq}. The two-forms satisfy the relations
\begin{equation}\label{relations_two_forms}
\Omega \wedge \bar{\Omega} = 2 J \wedge J \ne 0 \ , \qquad \Omega \wedge J = 0 \ , \qquad \Omega \wedge \Omega = 0 \ ,
\end{equation}
while the one-form is such that
\begin{equation}\label{K_compatible}
K \cdot K =0 \ , \qquad \bar{K} \cdot K = 2 \ , \qquad \iota_K J =0 \ , \qquad \iota_K \Omega = \iota_{\bar{K}} \Omega =0 \ .
\end{equation}
These forms can be expressed in terms of the spinors as follows,
\begin{align}
K_m & = \bar{\eta}^c_{2} \gamma_m \eta_{1} \ , \label{definition_one-form_K} \\
J_{mn} & = \tfrac{1}{2} \iu \left(\bar{\eta}_{1} \gamma_{mn} \eta_{1} + \bar{\eta}_{2} \gamma_{mn} \eta_{2}\right) \ ,\qquad \Omega_{mn} =  \bar{\eta}_{2} \gamma_{mn} \eta_{1} \ , \label{definition_two-forms}
\end{align}
where $\gamma_m$, $m=1,\ldots,6$, are $\so6$ gamma-matrices and $\gamma_{mn} = \frac12(\gamma_m \gamma_n - \gamma_n \gamma_m)$. By using Fierz identities and assuming that each $\eta^i$ satisfies $\bar\eta^i \eta^i=1$, it can be checked that these definitions for $K$, $J$ and $\Omega$ indeed fulfill the relations \eqref{relations_two_forms} and \eqref{K_compatible}.

The existence of the one-form~$K$ allows one to define an almost product structure~${P_m}^n$ on the manifold through the expression
\begin{equation}
\label{almost_product_structure}
{P_m}^n = K_m \bar{K}^n + \bar{K}_m K^n - \delta_m^{\phantom{m}n} \ .
\end{equation}
Using \eqref{K_compatible}, it is easy to check that~${P_m}^n$ does square to the identity, that is
\begin{equation}\label{P=1}
 P_m^{\phantom{m}n} P_n^{\phantom{n}p} = \delta_m^{\phantom{m}p} \ .
\end{equation}
From the definition~\eqref{almost_product_structure} and the first two relations in~\eqref{K_compatible}, it can be seen that $K_m$ and $\bar{K}_m$ are eigenvectors of $P_m^{\phantom{m}n}$ with eigenvalue $+1$. Also, all vectors simultaneously orthogonal to $K_m$ and $\bar{K}_m$ have eigenvalue $-1$. Thus $K_m$ and $\bar{K}_m$ span the $+1$ eigenspace and as a consequence the tangent space of $\M$ splits as
\begin{equation}\label{tangent_space_splitting}
T \M = T_2 \M \oplus T_4 \M \ ,
\end{equation}
where $T_2 \M$ has a trivial structure group and is spanned by $\Re\,K^m$ and $\Im\,K^m$. We can then choose a basis of one-forms $v^i$, $i=1,2$ on $T_2 \M$ normalized as
\begin{equation} \label{one-forms_relation}
 v^i \wedge v^j = \epsilon^{ij}\, \vol_2 \ ,
\end{equation}
where $\vol_2$ is the volume form on $T_2 \M$.

From the last constraints in~\eqref{K_compatible}, it follows that the
two-forms~$J$ and~$\Omega$ have `legs' only along~$T_4 \M$. The three real two-forms $J^1=\Re\,\Omega$, $J^2=\Im\,\Omega$ and $J^3=J$ form a triplet of symplectic two-forms on~$T_4 \M$ and from~\eqref{relations_two_forms} we infer that
\begin{equation}\label{j_wedge_j}
J^\alpha \wedge J^\beta = 2 \delta^{\alpha\beta} \vol_4 \ ,\qquad \alpha,\beta =1,2,3\ ,
\end{equation}
where $\vol_4$ denotes the volume form on~$T_4
\M$. Eq.~\eqref{j_wedge_j} states that the~$J^\alpha$ span a space-like
three-plane in the space of two-forms on~$T_4 \M$. The
triplet~$J^\alpha$ therefore defines an~$\su2$ structure on~$T_4
\M$. Finally, note that any pair of spinors $\tilde \eta^i$ which is
related to~$\eta^i$ by an~$\su2 \simeq \so3$ transformation defines
the same $\su2$ structure \cite{ReidEdwards:2008rd}. The one-form~$K$ is invariant under this rotation but the two-forms~$J^\alpha$ transform as a triplet.\footnote{Note also that the phase of $K$ corresponds to the overall phase of the pair $\eta^i$.} Thus there is an~$\su2$ freedom in the parametrization of the~$\su2$ structure. This $\su2$ is a subgroup of the R-symmetry group $\su4$ of $\mathcal{N}=4$ supergravity. 

The case when all forms $K$, $J$ and $\Omega$ (or equivalently $v^i$
and $J^\alpha$) are closed corresponds to a manifold $\M$ having
$\su2$ holonomy. This can be seen from
Eq.~\eqref{definition_one-form_K} and~\eqref{definition_two-forms},
since these forms being closed translates into the spinors $\eta^i$
being covariantly constant with respect to the Levi-Civita
connection. The only such manifold in six dimensions is the product
manifold $K3 \times T^2$, that is the product of a $K3$ manifold with a
two-torus. In that case, the almost product structure $P$ is trivially realized by the Cartesian product. 

\subsection{Kaluza-Klein data} \label{sec:KKdata}

So far, we analyzed the parametrization of an~$\su2$ structure over a single point of~$\M$. This gives all deformations of the~$\su2$ structure. But in order to find the low-energy effective action we have to perform a Kaluza-Klein truncation of the spectrum and thereby eliminate all modes with a mass above the compactification scale. This we do in two steps. First, we have to ensure that there are no massive gravitino multiplets in the $\mathcal{N}=4$ theory. It can be shown that these additional gravitino multiplets are~$\su2$ doublets which must therefore be projected out \cite{Grana:2005ny,Triendl:2009ap}. This also automatically removes all one- and three-forms in the space of forms acting on tangent vectors in~$T_4 \M$. Furthermore, the splitting~\eqref{tangent_space_splitting} becomes rigid, since a variation of this splitting is parametrized by a two-form with one leg on~$T_2 \M$ and the other on~$T_4 \M$ over each point of~$\M$, but one-forms acting on~$T_4 \M$ are projected out.

In the following, we will make the additional assumption that the almost product structure~\eqref{almost_product_structure} is integrable. This means that every neighborhood $\U$ of $\M$ can be written as a product $\U_2 \times \U_4$ such that $T_2 \M$ and $T_4 \M$ are tangent to $U_2$ and $U_4$, respectively. In other words, local coordinates $z^i,i=1,2$ and $y^a, a=1,\dotsc,4$ can be introduced on $\M$ such that $T_2 \M$ is generated by $\partial/\partial z^i$ and $T_4 \M$ by $\partial/\partial y^a$. The metric on~$\M$ can therefore be written in block-diagonal form as
\begin{equation}\label{metric}
\diff s^2 = g_{ij}(z,y)\, \diff z^i \diff z^j + g_{ab}(z,y)\, \diff y^a \diff y^b\ .
\end{equation}

In a second step, we truncate the infinite set of differential forms on $\M$ to a finite-dimensional subset. This chooses the light modes out of an infinite tower of (heavy) KK-states. This has to be done in a consistent way, \emph{i.e.}~such that only (but also all) scalars with masses below a chosen scale are kept in the low-energy spectrum.

Let us denote by $\Lambda^2 T_4 \M$ the space of two-forms on $\M$
that vanish identically when acting on tangent vectors in~$T_2
\M$. The Kaluza-Klein truncation means that we only need to consider
an $n$-dimensional subspace~$\Lambda_\mathrm{KK}^2 T_4 \M$ having
signature~$(3,n-3)$ with respect to the wedge product. The two-forms
$J^\alpha$ span a space-like three-plane in~$\Lambda_\mathrm{KK}^2 T_4
\M$ and therefore parametrize the space \cite{Triendl:2009ap}
\begin{equation}\label{moduli_space}
\mathcal{M}_{J^\alpha} = \frac{\so{3,n-3}}{\so3\times\so{n-3}}
\end{equation}
with dimension $3n-9$. Together with the volume~$\vol_4 \sim \e^{-\rho}$ this gives~$3n-8$ geometric scalar fields on~$T_4 \M$. Let us choose a basis $\omega^I$, $I=1,\dots,n$ on $\Lambda_\mathrm{KK}^2 T_4 \M$ such that
\begin{equation}\label{omega_wedge_omega}
\omega^I \wedge \omega^J = \eta^{IJ} \e^{\rho} \vol_4 \ ,
\end{equation}
with $\eta^{IJ}$ being the (symmetric) intersection matrix with
signature~$(3,n-3)$. The factor~$\e^\rho$ was introduced in order
to keep $\omega^I$ and~$\eta^{IJ}$ independent of the  volume modulus.

The remaining geometric scalars are parametrized by $K$. The latter is a complex one-form acting on~$T_2 \M$ which can be expanded in terms of the~$v^i$ fulfilling eq.~\eqref{one-forms_relation}. The overall real factor of~$K$ is proportional to the square root of~$\vol_2$, while the overall phase of~$K$ is not physical.\footnote{The overall phase of $K$ corresponds to the overall phase of the spinor pair $\eta^i$, which is of no physical relevance.} The other two degrees of freedom in $K$ parametrize the complex structure on~$T_2 \M$. This gives altogether three geometric scalars on~$T_2 \M$.

On a generic manifold with $\su2$ structure, the one- and two-forms are not necessarily closed. On the truncated subspace we just introduced, one can generically have\cite{Spanjaard:2008zz,ReidEdwards:2008rd}
\begin{equation}\label{differential_algebra}
\begin{aligned}
\diff v^i &= \tor^i v^1\wedge v^2+ \torr^i_I \omega^I\ , \\
\diff \omega^I &= {\tilde\Tor}_{iJ}^I v^i \wedge \omega^J\ ,
\end{aligned}
\end{equation}
where the parameters $\tor^i$, $\torr^i_I$ and ${\tilde\Tor}_{iJ}^I$ are constant. Indeed, eqs.~\eqref{differential_algebra} state that $J^\alpha$ and $K$ are in general not closed, their differential being related to the torsion classes of the manifold\cite{Gauntlett:2003cy}. The parameters in the r.h.s.~of~\eqref{differential_algebra} play the role of gauge charges in the low-energy effective supergravity, as we will see in section~\ref{subsec:NS}.

One can show that demanding integrability of the almost product structure~\eqref{almost_product_structure} forces~$\torr^i_I$ to vanish\cite{Louis:2009dq}. The reason is that in such a case it is impossible to generate a form in $\Lambda^2 T_4 \M$ like $\omega^I$ by differentiating a one-form $v^i$ that acts non-trivially only on vectors in $T_2 \M$. We will therefore restrict the discussion in the following to this case and set $t^i_I=0$.

On the other hand, the parameters~$\tor^i$ and~$\tilde\Tor^I_{iJ}$ are not completely arbitrary but constrained by Stokes' theorem and nilpotency of the $\diff$-operator. Acting with $\diff$ on eqs.~\eqref{differential_algebra} and using $\diff^2=0$ leads to
\begin{equation}\label{constraint_quadratic}
\tor^i {\tilde\Tor}^I_{iJ} - \epsilon^{ij} {\tilde\Tor}^I_{iK} {\tilde\Tor}^K_{jJ} = 0 \ ,
\end{equation}
where we choose $\epsilon^{12}=1$.
On the other hand, Stokes' theorem implies the vanishing of $\int_Y \diff (v^i \wedge\omega^I\wedge\omega^J)$ for any compact $\M$, which yields
\begin{equation}\label{constraint_linear}
\tor^i \eta^{IJ} - \epsilon^{ij} {\tilde\Tor}^I_{jK} \eta^{KJ}  - \epsilon^{ij} {\tilde\Tor}^J_{jK} \eta^{KI} = 0 \ .
\end{equation}
This in turn implies that ${\tilde\Tor}^I_{iJ}$ can be written as
\begin{equation}\label{expression_torsion_matrix}
{\tilde\Tor}^I_{iJ} = \tfrac12\, \epsilon_{ij} \tor^j \delta^I_J + \Tor^I_{iJ} \ ,
\end{equation}
with $\epsilon_{12} = -1$ and $\Tor^I_{iJ}$ satisfying
\begin{equation}\label{so3n}
\Tor^I_{iK} \eta^{KJ} = - \Tor^J_{iK} \eta^{KI}\ .
\end{equation}
It will be useful to define two $n \times n$ matrices $\Tor_i = (\Tor_{i})_J^I$, which due to \eqref{so3n} are in the algebra of $\so{3,n-3}$. Finally, substituting $\torr^i_I = 0$ and~\eqref{expression_torsion_matrix} into~the expressions~\eqref{differential_algebra} we are left with
\begin{equation}\label{differential_algebra_simpler}
\begin{aligned}
\diff v^i &= \tor^i v^1\wedge v^2 \ , \\
\diff \omega^I &= \tfrac12\, \tor^i \epsilon_{ij} v^j \wedge \omega^I + \Tor_{iJ}^I v^i \wedge \omega^J\ ,
\end{aligned}
\end{equation}
where, according to eq.~\eqref{constraint_quadratic}, the matrices $\Tor_i$ satisfy the commutation relation
\begin{equation}\label{XXXX}
[\Tor_1, \Tor_2] = \tor^i \Tor_i \ .
\end{equation}

If all parameters $\tor^i$ and $\Tor^I_{iJ}$ vanish, we recover the case with closed forms $v^i$ and $J^\alpha$ and consequently the manifold is $K3 \times T^2$. In this case, the two-forms $\omega^I$ are harmonic and span the second cohomology of $K3$, their number being fixed to $n = 22$.

\section{The low-energy effective action} \label{sec:eff_action}

\subsection{The NS-NS sector} \label{subsec:NS}

As already mentioned in the introduction, the reduction of the NS-NS
sector is completely similar to that performed in
Ref.~\cite{Louis:2009dq} for the heterotic string, therefore we will
essentially only recall the results.

The massless fields arising from the NS-NS sector in type IIA
supergravity are the metric $g_{MN}$, the two-form $\mathcal{B}_2$ and
the dilaton $\Phi$. The ten-dimensional action governing the dynamics of these fields is given by
\begin{equation}\label{action_NS_10}
S_\mathrm{NS} = \tfrac12 \int_{M_{1,3} \times \M}  \e^{-2\Phi} \big( \mathcal{R} + 4 \diff \Phi \wedge \ast \diff \Phi - \tfrac12 \mathcal{H}_3 \wedge \ast \mathcal{H}_3 \big) \ ,
\end{equation}
where $\mathcal{R}$ is the Ricci scalar and $\mathcal{H}_3=\diff \mathcal{B}_2$ is the field-strength of the two-form~$\mathcal{B}_2$. A KK ansatz for these fields can be written as
\begin{equation}\label{KK_expansion_NS}
\begin{aligned}
\diff s^2 &= g_{\mu\nu} \diff x^\mu \diff x^\nu + g_{ij} \mathcal{E}^i \mathcal{E}^i + g_{ab} \diff y^a \diff y^b \ , \\
\mathcal{B}_2 &= B + B_i \wedge \mathcal{E}^i  + b_{12} \mathcal{E}^1 \wedge \mathcal{E}^2 + b_I \omega^I \ ,
\end{aligned}
\end{equation}
where we have defined the `gauge-invariant' one-forms 
$\mathcal{E}^i = v^i - G^i_\mu \diff x^\mu$. 
The expansion of the ten-dimensional
two-form~$\mathcal{B}_2$ leads to a set of four-dimensional fields: a
two-form~$B$, two vectors or one-forms~$B_i$ and~$n+1$ scalar
fields~$b_I$ and $b_{12}$.\footnote{Note that in this paper we do not
  consider background flux for $\mathcal{H}_3$. This situation 
  has been discussed for example in
  \cite{Aldazabal:2008zza,Dall'Agata:2009gv,Dibitetto:2010rg}
   where it was shown that, as usual, the background fluxes appear as 
   gauge charges in the effective action which gauge specific directions 
   in the $N=4$ field space.}
In computing the low-energy effective action, one has to express
the variation of the metric components~$g_{ab}$ in terms of the~$3n-8$
geometric moduli on $T_4 \M$ or, more precisely, one needs an expression for the
line element $g^{ac} g^{bd} \delta g_{ab} \delta g_{cd}$. As
a first step one expands the two-forms~$J^\alpha$ parametrizing the
$\su2$ structure in terms of the basis $\omega^I$ according to
\begin{equation}\label{expansion_j}
J^\alpha = \e^{-\frac\rho2} \zeta^\alpha_I \omega^I \ .
\end{equation}
However, the~$3n$ parameters~$\zeta^\alpha_I$ are not all
independent. Inserting the expansion~\eqref{expansion_j} into
Eq.~\eqref{j_wedge_j}, and using the
relation~\eqref{omega_wedge_omega}, one obtains the six independent constraints
\begin{equation}
\eta^{IJ} \zeta^\alpha_I \zeta^\beta_J = 2 \delta^{\alpha\beta} \ .
\end{equation}
Moreover, an~$\so3$ rotation acting on the upper index
of~$\zeta^\alpha_I$ gives new two-forms~$J^\alpha$ that are linear
combinations of the old ones, defining therefore the same three-plane
and leaving us at the same point of the moduli space. Altogether, we
end up with the right number of $3n-9$ geometric moduli
parametrizing~$\mathcal{M}_{J^\alpha}$ in Eq.~\eqref{moduli_space}. Furthermore,
Ref.~\cite{Louis:2009dq} derived the line element to be
\begin{equation}\label{line_element}
g^{ac} g^{bd} \delta g_{ab} \delta g_{cd} = \delta\rho^2 + (2\eta^{IJ} - \zeta^{\alpha I} \zeta^{\beta J}) \delta\zeta^\alpha_I \delta\zeta^\beta_J \ ,
\end{equation}
where~$\zeta^{\alpha I} = \eta^{IJ} \zeta^\alpha_J$. Note that this
expression is indeed the metric on the coset
\begin{equation}
\mathbb{R}^+ \times \frac{\so{3,n-3}}{\so3 \times \so{n-3}} \ .
\end{equation} 

With the last result at hand, it is straightforward to insert the ansatz~\eqref{KK_expansion_NS} into the action~\eqref{action_NS_10} and obtain the effective four-dimensional action
\begin{equation}\label{action_NS}
\begin{aligned}
S_\mathrm{NS} = \tfrac12 \int_{M_{1,3}} & \Big[ R \ast 1 - \tfrac12 \e^{-4\phi} \big\vert \Dc B \big\vert^2 - \tfrac12 \e^{-2\phi-\eta} \tilde{g}_{ij} \Dc G^i \wedge \ast \Dc G^j \\
& - \tfrac12 \e^{-2\phi+\eta} \tilde{g}^{ij} \big( \Dc B_i - b_{12} \epsilon_{ik} \Dc G^k \big) \wedge \ast \big( \Dc B_j - b_{12} \epsilon_{jl} \Dc G^l \big) \\
& - \vert \diff \phi \vert^2 - \tfrac12 \e^{2\eta} \big( \vert \Dc b_{12} \vert^2 + \vert \Dc \e^{-\eta} \vert^2 \big) - \tfrac14 \tilde{g}^{ik} \tilde{g}^{jl} \Dc \tilde{g}_{ij} \wedge \ast \Dc \tilde{g}_{kl} \\
& - \tfrac14 \vert \Dc\rho \vert^2 - \tfrac14 (H^{IJ} - \eta^{IJ}) \Dc\zeta^\alpha_I \wedge \ast \Dc\zeta^\beta_J - \tfrac12 \e^\rho H^{IJ} \Dc b_I \wedge \ast \Dc b_J \\
& - \tfrac54 \e^{2\phi+\eta} \tilde{g}_{ij} \tor^i \tor^j + \tfrac18 \e^{2\phi+\eta} \tilde{g}^{ij} {[H, \Tor_i]^I}_J {[H, \Tor_j]^J}_I \\
& - \tfrac18 \e^{2\phi-\eta+\rho} \tilde{g}_{ij} \tor^i \tor^j H^{IJ} b_I b_J - \tfrac12 \e^{2\phi+\eta+\rho} \tilde{g}^{ij} H^{IJ} \Tor^K_{iI} \Tor^L_{jJ} b_K b_L \Big] \ ,
\end{aligned}
\end{equation}
where $R$ denotes the Ricci scalar in four-dimensions and we have introduced the notation $\vert f \vert^2 = f \wedge \ast f$ for any form~$f$. Moreover, the symmetric matrix~$H^{IJ}$ is defined according to $\omega^I \wedge \ast \omega^J = H^{IJ} \e^\rho \vol_4$, which can be expressed in terms of the parameters~$\zeta^\alpha_I$ by \cite{Louis:2009dq}\footnote{This expression can be derived by using the fact that the two-forms $J^\alpha$ are self-dual, $J^\alpha = \ast J^\alpha$, with all other orthogonal linear combinations of the $\omega^I$ being anti-self dual.}
\begin{equation}
H^{IJ} = - \eta^{IJ} + \zeta^{\alpha I} \zeta^{\alpha J} \ .
\end{equation}
(The commutators in \eqref{action_NS} use ${H^I}_J = H^{IK}
\eta_{KJ}$.) In the two-dimensional metric $g_{ij}$ defined in
\eqref{metric} we separated the overall volume $\e^{-\eta}$ from the
other two independent (complex structure) degrees of freedom by
introducing the rescaled metric~$\tilde{g}_{ij} = \e^\eta g_{ij}$. 
It satisfies $\det \tilde{g} = 1$ and can be expressed in terms of a complex-structure parameter~$\kappa$ as
\begin{equation}
\tilde{g}_{ij} = \frac1{\Im \kappa} \begin{pmatrix}
1 & \Re \kappa \\ \Re \kappa & \vert \kappa \vert^2
\end{pmatrix} \ .
\end{equation}
In order to write the action in the Einstein frame, we also performed
the Weyl rescaling~$g_{\mu\nu} \to \e^{2\phi} g_{\mu\nu}$ of the
four-dimensional metric, where $\phi = \Phi + \frac12 (\eta + \rho)$
is the four-dimensional dilaton. Finally, the various non-Abelian
field-strengths and covariant derivatives in \eqref{metric} are given
by
\begin{subequations}\label{fieldstrengths_NS}
\begin{align}
\Dc B & = \diff B + B_i \wedge \Dc G^i \ , \\
\Dc G^i & = \diff G^i - \tor^i G^1 \wedge G^2 \ , \label{DGi} \\
\Dc B_i & = \diff B_i + \epsilon_{ij} \tor^k  G^j \wedge B_k \ , \\
\Dc \tilde{g}_{ij} & = \diff \tilde{g}_{ij} + (\epsilon_{il} \tilde{g}_{jk} + \epsilon_{jl} \tilde{g}_{ik} - \epsilon_{kl} \tilde{g}_{ij}) \tor^k G^l \ , \\
\Dc \e^{-\eta} & = \diff \e^{-\eta} - \epsilon_{ij} t^j \e^{-\eta} G^i \ , \label{Deta} \\
\Dc b_{12} & = \diff b_{12} - \epsilon_{ij} \tor^j b_{12} G^i - \tor^i B_i \ , \label{Db12} \\
\Dc\rho & = \diff\rho - \epsilon_{ij} \tor^j G^i \ , \\
\Dc\zeta^\alpha_I & = \diff\zeta^\alpha_I + \Tor^J_{iI} \zeta^\alpha_J G^i \ , \\
\Dc b_I & = \diff b_I + \tilde{\Tor}^J_{iI} b_J G^i \ .
\end{align}
\end{subequations}
As a next step let us  turn to the R-R sector.

\subsection{The R-R sector}\label{subsec:RR}

So far, we have reduced the kinetic term for the NS fields. The
remaining part of the ten-dimensional action for type IIA supergravity consists of the kinetic terms for the R-R fields and the Chern-Simons term,
\begin{align}\label{action_RR_10}
S_\mathrm{RR} & = -\tfrac14 \int_{ M_{1,3} \times \M } \big(\mathcal{F}_2\wedge \ast \mathcal{F}_2 + \tilde{\mathcal{F}}_4 \wedge \ast \tilde{\mathcal{F}}_4 \big) \ , \\
S_\mathrm{CS} & = - \tfrac14 \int_{ M_{1,3} \times \M } \mathcal{B}_2\wedge \mathcal{F}_4 \wedge \mathcal{F}_4 \ ,\label{action_CS_10}
\end{align}
where $\mathcal{F}_2 = \diff \mathcal{A}_1$ and $\mathcal{F}_4 = \diff
\mathcal{C}_3$.  $\tilde{\mathcal{F}}_4$ is the modified field
strength of $\mathcal{C}_3$ defined as
\begin{equation}\label{modified_fieldstrength}
\tilde{\mathcal{F}}_4= \diff \mathcal{C}_3 - \mathcal{A}_1 \wedge \diff \mathcal{B}_2.
\end{equation}
Analogously to the KK ansatz~\eqref{KK_expansion_NS}, we expand the ten-dimensional RR fields in the set of internal one-forms~$\Ei$ and two-forms~$\omega^I$ as follows,
\begin{equation}\label{KK_expansion_RR}
\begin{aligned}
\mathcal{A}_1 = & {} A + a_i \Ei \ , \\
\mathcal{C}_3 = & {} (C - A \wedge B) + (C_i - A \wedge B_i) \wedge \Ei \\
& {} + (C_{12} - b_{12} A) \wedge \Eone \wedge \Etwo + (C_I - b_I A) \wedge \omega^I + c_{iI} \Ei \wedge \omega^I \ .
\end{aligned}
\end{equation}
In terms of four-dimensional fields we thus have a three-form~$C$, two two-forms~$C_i$, $2+n$ vectors or one-forms~$A$, $C_{12}$ and $C_I$, and finally $2n+2$ scalars~$a_i$ and~$c_{iI}$.\footnote{As for the $B$-field, we also do not 
consider background fluxes for the RR field strengths in this paper.
Their effect is similar to an $\mathcal{H}_3$ flux in that additional
directions    in the $N=4$ field space become gauged
  \cite{Aldazabal:2008zza,Dall'Agata:2009gv,Dibitetto:2010rg}.} 
In the expansion of the three form $\mathcal{C}_3$, it is convenient to introduce some mixing with the four-dimensional components from $\mathcal{A}_1$ and $\mathcal{B}_2$. The reason for this is that in this case the four-dimensional field strengths $\diff C$, $\diff C_i$, $\diff C_{12}$ and $\diff C_I$ remain invariant under the gauge transformations
\begin{equation}\label{pformgaugetransf}
\begin{aligned}
\mathcal{A}_1 & \to \mathcal{A}_1 + \diff \Lambda\ , \\
\mathcal{B}_2 & \to \mathcal{B}_2 + \diff \Lambda_1\ ,\\
\mathcal{C}_3 & \to \mathcal{C}_3 + \diff \Lambda_2 + \Lambda \diff \mathcal{B}_2\ ,
\end{aligned}
\end{equation}
which is a symmetry of type IIA supergravity, as can be seen from the modified field-strength~\eqref{modified_fieldstrength}.

Before we continue, let us pause and count the total number of light
modes arising from the KK ansatz in the NS-NS plus RR-sector. From
Eq.~\eqref{KK_expansion_NS} (and the subsequent analysis) we learn
that the spectrum  in the NS-sector 
contains the graviton, a two-form $B$, four vectors
$G^i, B_i$ and $4n-3$ scalars. From Eq.~\eqref{KK_expansion_RR}, we
see that two two-forms, $2+n$ vectors and $2n+2$ scalars arise in the
RR-sector. After dualizing the three two-forms to scalars we thus have
a total spectrum of a graviton,  $6+n$ vectors and $6n+2$ scalars. As
we review in the next section, this is indeed the spectrum of an $\mathcal{N}=4$ supergravity with $n$ vector multiplets.

Substituting this expansion for the ten-dimensional fields into the action~\eqref{action_RR_10} and performing at the end the Weyl rescaling~$g_{\mu\nu} \to \e^{2\phi} g_{\mu\nu}$, we obtain 
\begin{equation}\label{action_RR_kin}
\begin{aligned}
S_\mathrm{RR}= -\tfrac14 \int_{M_{1,3}} & \Big[ \e^{-\eta-\rho} \big\vert \diff A - a_i \Dc G^i \big\vert^2 + \e^{-4\phi-\eta-\rho} \big\vert \Dc C - \diff A \wedge B \big\vert^2 \\
& + \e^{-2\phi-\rho} \tilde{g}^{ij} \big( \Dc C_i - \diff A \wedge B_i + a_i \Dc B \big) \wedge \\
& \hspace{2in} \wedge \ast \big( \Dc C_j - \diff A \wedge B_j + a_j \Dc B \big) \\
& + \e^{\eta-\rho} \big\vert \Dc C_{12} - b_{12} \diff A - a_i (\epsilon^{ij} \Dc B_j - b_{12} \Dc G^i) \big\vert^2 \\
& + \e^{-\eta} H_{IJ} \big( \Dc C^I - b^I \diff A - c_i^I \Dc G^i \big) \wedge \ast \big( \Dc C^J - b^J \diff A - c_j^J \Dc G^j \big) \\
& + \e^{2\phi} \tilde{g}^{ij} H_{IJ} \big( \Dc c_i^I + a_i \Dc b^I \big) \wedge \ast \big( \Dc c_j^J + a_j \Dc b^J \big) \\
& + \e^{2\phi-\rho} \tilde{g}^{ij} \Dc a_i \wedge \ast \Dc a_j + \e^{4\phi+\eta-\rho} (\tor^i a_i)^2 \ast 1 \\
& + \e^{4\phi + \eta} H_{IJ} \big[\epsilon^{ij} \Tor^I_{iK} (c_j^K + a_j b^K) - \tor^i (c_i^I - a_i b^I) \big] \cdot \\
& \hspace{1.5in} \cdot \big[\epsilon^{kl} \Tor^J_{kL} (c_l^L + a_l b^L) - \tor^k (c_k^J - a_k b^J) \big] \ast 1 \Big] \ .
\end{aligned}
\end{equation}
On the other hand, the Chern-Simons term \eqref{action_CS_10}
gives the following contribution
\begin{equation}\begin{aligned}\label{action_RR_CS}
S_\mathrm{CS} = - \tfrac14 \int_{M_{1,3}} & \Big[ \, 2 \epsilon^{ij} c^J_i \tilde{\Tor}^I_{jJ} b_I \big( \Dc C - \diff A \wedge B \big) \\
& - 2 \big(\Dc C_i - \diff A \wedge B_i \big) \wedge \epsilon^{ij} b_I \Dc c_j^I + b_{12} \eta_{IJ} \Dc C^I \wedge \Dc C^J \\
& + 2 \big( \Dc C_{12} - b_{12} \diff A \big) \wedge b_I \big(\Dc C^I - \tfrac12 b^I \diff A - c_i^I \Dc G^i \big) \\
& - \Dc B \wedge \epsilon^{ij} c_{iI} \big( \Dc c_j^I - \tilde{\Tor}_{jJ}{}^I C^J \big) + 2 B_i \wedge \epsilon^{ij} \tilde{\Tor}_{jIJ} C^I \wedge \Dc C^J \\
& - 2 \big( \Dc B_i - b_{12} \epsilon_{ik} \Dc G^k \big) \wedge \epsilon^{ij} c_{jI} \big( \Dc C^I - \tfrac12 c_l^I \Dc G^l \big) \Big] \ .
\end{aligned}\end{equation}
The non-Abelian field-strengths and covariant derivatives of all
four-dimensional RR-fields are given by
\begin{subequations}\label{fieldstrengths_RR}
\begin{align}
\Dc C & = \diff C - C_i \wedge \Dc G^i \ , \\
\Dc C_i & = \diff C_i + \epsilon_{ij} \tor^k G^j \wedge C_k + \epsilon_{ij} C_{12} \wedge \Dc G^j \ , \\
\Dc C_{12} & = \diff C_{12} + t^i C_i - \epsilon_{ij} \tor^j G^i \wedge C_{12} \ , \label{DC12} \\
\Dc C^I & = \diff C^I + \tilde{\Tor}_{iJ}{}^I G^i \wedge C^J\ ,  \label{DCI} \\
\Dc a_i & = \diff a_i + \epsilon_{ij} \tor^k a_k G^j \ , \\
\Dc c_i^I & = \diff c_i^I + \epsilon_{ij} \tor^k c_k^I G^j - \tilde{\Tor}_{jJ}{}^I c_i^J G^j + \tilde{\Tor}_{jJ}{}^I C^J \ .
\end{align}
\end{subequations}

Let us summarize. The bosonic part of the low-energy four-dimensional effective action arising from the compactification of type IIA supergravity on~$\su2$-structure manifolds is given by the sum of the contribution from the NS-NS sector, Eq.~\eqref{action_NS}, and the contribution from the RR sector, Eqs.~\eqref{action_RR_kin} and~\eqref{action_RR_CS}, that is
\begin{equation}\label{totalaction}
S_\mathrm{eff} = S_\mathrm{NS} + S_\mathrm{RR} + S_\mathrm{CS} \ .
\end{equation}
The covariant derivatives and field strengths corresponding to the various four-dimensional fields are given in Eqs.~\eqref{fieldstrengths_NS} and~\eqref{fieldstrengths_RR}.

The next step is  to establish the consistency of this action with four-dimensional $\mathcal{N}=4$ supergravity. To do this, we will bring the action into the canonical form proposed in Ref.~\cite{Schon:2006kz} by performing a series of field redefinitions.

\section{Consistency with $\mathcal{N}=4$ supergravity} \label{sec:ConsistencyN=4}

The gravity multiplet of $\mathcal{N}=4$ supergravity in four
dimensions contains as bosonic degrees of freedom the metric, six massless vectors and two real
scalars while a vector multiplet consist of a massless vector field
and six real scalars.  $\mathcal{N}=4$ supergravity coupled to $n$
vector multiplets has a global symmetry $\mathrm{SL}(2) \times
\so{6,n}$ and the scalar fields of the theory assemble into a complex
field $\tau$ describing an $\mathrm{SL}(2)/\so2$ coset and a
$(6+n)\times(6+n)$ matrix $M_{MN}$ parametrizing the coset
\begin{equation}
\frac{\mathrm{SO}(6,n)}{\mathrm{SO}(6)\times \mathrm{SO}(n)} \ .
\end{equation}

In Ref.~\cite{Schon:2006kz}, the action of the  most general
gauged $\mathcal{N}=4$ supergravity is given using the embedding
tensor formalism. 
All possible gaugings are encoded in two tensors,~$f_{\alpha MNP}$
and~$\xi_{\alpha M}$, where $\alpha$ is an~$\mathrm{SL}(2)$ index
taking the values $+$ and $-$. As it turns out, for the effective
action \eqref{totalaction}
both $f_{-MNP}$ and $\xi_{-M}$ vanish, and therefore we choose to
start with the formulas of Ref.~\cite{Schon:2006kz} adapted to this
case. In order to simplify the notation, we omit the $\alpha=+$ index
in the couplings~$f_{+MNP}$ and~$\xi_{+M}$ and write simply $f_{MNP}$
and $\xi_M$ for the non-trivial couplings. With this in mind, the action for gauged $\mathcal{N}=4$ supergravity can be divided in three parts,
\begin{equation}\label{N=4general}
S_{\mathcal{N}=4} = S_\mathrm{kin} + S_\mathrm{top} + S_\mathrm{pot} \ ,
\end{equation}
that is kinetic, topological and potential terms. The part of the action containing the kinetic terms reads
\begin{multline}\label{N=4_canonical_kin}
S_\mathrm{kin} = \tfrac12 \int_{M_{1,3}} \big[ R \ast 1 + \tfrac18 \Dc M_{MN} \wedge \ast \Dc M^{MN} - \tfrac12 (\Im \tau)^{-2} \Dc \tau \wedge \ast \Dc \bar\tau \\
{} - (\Im \tau) \, M_{MN} \Dc V^{M+} \wedge \ast \Dc V^{N+} + (\Re \tau) \, \eta_{MN} \Dc V^{+} \wedge \Dc V^{N+} \big] \ ,
\end{multline}
where the constant matrix $\eta_{MN}$ is an $\mathrm{SO}(6,n)$ metric and the non-Abelian field-strengths for the electric vector fields $V^{M+}$ are given by the expression
\begin{equation}\label{fieldstrengths_embedding_tensor}
\Dc V^{M+} = \diff V^{M+} - \tfrac12 {\hat{f}}_{NP}{}^M V^{N+} \wedge V^{P+} + \tfrac12 \xi^M B^{++} \ , 
\end{equation}
where  $B^{++}$ is an auxiliary two-form whose role we soon
explain.\footnote{As noted above, we omit the $+$ index of
  Ref.~\cite{Schon:2006kz} in the couplings $f_{MNP}$ and~$\xi_{M}$,
  but we do keep it for the gauge fields and denote the electric
  vectors by $V^{M+}$ while the magnetic vectors are $V^{M-}$.} 
The covariant derivatives of the scalar fields are defined as
\begin{align}
\Dc \tau & = \diff \tau + \xi_M \tau V^{M+} + \xi_M V^{M-} \ , \label{Dtau} \\
\Dc M_{MN} & = \diff M_{MN} + \Theta_{PM}{}^Q M_{NQ} V^{P+} + \Theta_{PN}{}^Q M_{MQ} V^{P+} \ .
\end{align}
In these expressions, the following useful shorthands were used,
\begin{align}
\hat{f}_{MNP} & = f_{MNP} - \tfrac12 \xi_M \eta_{PN} + \tfrac12 \xi_P \eta_{MN} - \tfrac32 \xi_N \eta_{MP} \ , \\
\Theta_{MNP} & = f_{MNP} - \tfrac12 \xi_N \eta_{PM} - \tfrac12 \xi_P \eta_{NM} \ .
\end{align}

As we can see, the presence of an auxiliary two-form field~$B^{++}$ is
related to the fact that the complex scalar~$\tau$ is charged with
respect to the magnetic duals~$V^{M-}$ of the electric vector
fields~$V^{M+}$. The two-form~$B^{++}$ 
acts as a Lagrange multiplier, in the sense that its equation of motion merely ensures that~$V^{M-}$ and~$V^{M+}$ are related by an electric-magnetic duality. This follows from the last term in the topological part of the $\mathcal{N}=4$ supergravity action
\begin{equation}\label{N=4top}
\begin{aligned}
S_\mathrm{top} = -\tfrac12 \int_{M_{1,3}} & \big[ \xi_M \eta_{NP} V^{M-} \wedge V^{N+} \wedge \diff V^{P+} 
- \tfrac14 \hat{f}_{MNR} \hat{f}_{PQ}{}^R V^{M+} \wedge V^{N+} \wedge V^{P+} \wedge V^{Q-} \\
&\quad - \xi_M B^{++} \wedge \big( \diff V^{M-} - \tfrac12 \hat{f}_{QR}{}^M V^{Q+} \wedge V^{R-} \big) \big] \ .
\end{aligned}
\end{equation}

Finally, there is also a potential energy that contributes to the action as
\begin{equation}\label{N=4pot}
\begin{aligned}
S_\mathrm{pot} = - \tfrac1{16} \int_{M_{1,3}} & (\Im \tau)^{-1} \big[ 3 \xi^M \xi^N M_{MN} \\
& + f_{MNP} f_{QRS} \big( \tfrac13 M^{MQ} M^{NR} M^{PS} + (\tfrac23 \eta^{MQ} - M^{MQ}) \eta^{NR} \eta^{PS} \big) \big] \ .
\end{aligned}
\end{equation}

\subsection{Field dualizations} \label{sec:dualizations}

The action $S_\mathrm{eff}$ that was obtained in \eqref{totalaction} does not have the same structure as the action given in Eq.~\eqref{N=4general}. Most obviously, the spectrum currently contains two-form fields, which we must replace by their dual scalar fields. Furthermore, as can be easily verified, the quadratic couplings of the vector field-strengths are not of the simple form seen in Eq.~\eqref{N=4_canonical_kin}, which implies that also some of the vector fields must be traded for their dual fields.

Our strategy will be the following. First we remove the
(non-dynamical) three-form field~$C$ from the theory and dualize the
two-forms~$B$ and~$C_i$ to scalars~$\beta$ and~$\gamma^i$,
respectively. In a second step, we determine the correct
electric-magnetic duality frame in which the action for the vector
fields takes the form \eqref{N=4_canonical_kin}. This we can do by
setting to zero the parameters~$\Tor^I_{iJ}$ and~$\tor^i$ determining
the charges, which makes it easier to perform electric-magnetic
duality transformations on the vector fields.  Once we have identified
the correct electric-magnetic duality frame, we can read off
the~$\so{6,n}$ coset matrix~$M_{MN}$, the complex scalar~$\tau$ and
the metric~$\eta_{MN}$. The final step is then to turn on the charges 
and use the information obtained in the previous steps to determine the components of the embedding tensor. Using the embedding tensor, we can then find the full expressions for the electric field strengths in the canonical action~\eqref{N=4_canonical_kin}, as well as the correct topological terms~\eqref{N=4top}. We can then verify that the action obtained in this way is equivalent to~$S_\mathrm{eff}$ by elimination of the extra two-form~$B^{++}$ introduced by the embedding tensor formalism.

As already mentioned, the four-dimensional three-form~$C$ carries no
degrees of freedom.  We can integrate it out using its equation of
motion. 
{}From the part of the effective action $S_\mathrm{eff}$ that depends on~$C$, namely
\begin{equation}\label{action_threeform}
S_{C} = - \tfrac14 \int_{M_{1,3}} \Big[ \e^{-4\phi-\eta-\rho} \big\vert \Dc C - \diff A \wedge B \big\vert^2 - 2 \epsilon^{ij} b_I \tilde{\Tor}^I_{iJ} c_j^J \big( \Dc C - \diff A \wedge B \big) \Big] \ ,
\end{equation}
follows the equation of motion
\begin{equation}\label{eqmotion_threeform}
\Dc C - \diff A \wedge B = - e^{4\phi+\eta+\rho} \epsilon^{ij} b_I \tilde{\Tor}^I_{iJ} c_j^J \ast 1 \ .
\end{equation}
Substituting this back into the action \eqref{action_threeform}, we obtain the potential term
\begin{equation}\label{potential_threeform}
S^\prime_C= - \tfrac14 \int_{M_{1,3}} \e^{4\phi+\eta+\rho} \big(
\epsilon^{ij} b_I \tilde{\Tor}^I_{iJ} c_j^J \big)^2 \ast 1\ .
\end{equation}

Next, we trade the two-forms~$C_i$ and~$B$ for their dual scalars. In contrast to the three-form~$C$, the two-forms~$C_i$ do not appear in the Lagrangian exclusively in the form~$\diff C_i$. As can be seen in the expression~\eqref{DC12} for the covariant field strength~$\Dc C_{12}$, they are also present as a St\"uckelberg-like mass term~$\tor^i C_i$, making it necessary to dualize the vector field~$C_{12}$ as well. Therefore, we dualize the~$C_i$ into scalar fields~$\gamma^i$ while at the same time dualizing the vector field~$C_{12}$ to a vector field~$\tilde{C}$. As already mentioned, the scalar field dual to $B$ will be called $\beta$. We present the details of this calculation in Appendix~\ref{sec:dualizations-appendix}.

After these steps, we arrive at an action~$S^\prime_\mathrm{eff}$ containing only scalar and vector fields (apart from the metric). The total action can be split into three components
\begin{equation}\label{Seff}
S^\prime_\mathrm{eff} = S_{\mathrm{scalar}} + S_{\mathrm{vector}} + S_{\mathrm{potential}}\,,
\end{equation}
where the kinetic terms for the scalar fields (and the four-dimensional metric) are
\begin{equation}\label{n=4scalar}
\begin{aligned}
S_{\mathrm{scalar}} = -\tfrac12 \int_{M_{1,3}} & \Big[ R \ast 1 + \vert \diff \phi \vert^2 + \tfrac12 \e^{2\eta} \big( \vert \Dc b_{12} \vert^2 + \vert \Dc \e^{-\eta} \vert^2 \big) + \tfrac14 \tilde{g}^{ik} \tilde{g}^{jl} \Dc \tilde{g}_{ij} \wedge \ast \Dc \tilde{g}_{kl} \\
& + \tfrac14 \vert \Dc\rho \vert^2 + \tfrac14 (H^{IJ} - \eta^{IJ}) \Dc\zeta^\alpha_I \wedge \ast \Dc\zeta^\beta_J + \tfrac12 \e^\rho H^{IJ} \Dc b_I \wedge \ast \Dc b_J \\
& + \e^{2\phi-\rho} \tilde{g}^{ij} \Dc a_i \wedge \ast \Dc a_j + \e^{2\phi} \tilde{g}^{ij} H_{IJ} (\Dc c_i^I + a_i \Dc b^I) \wedge \ast (\Dc c_j^J + a_j \Dc b^J) \\
& + \e^{2\phi+\rho} \tilde{g}^{ij} (\Dc \gamma_i + b^I \Dc c_{iI}) \wedge \ast (\Dc \gamma_j + b^J D c_{jJ}) \\
& + \e^{4\phi} \big\vert \Dc \beta - \epsilon^{ij} (a_i \Dc \gamma_j + a_i b_I \Dc c_j^I - \tfrac12 c_{iI} \Dc c_j^I) \big\vert^2 \Big] \ .
\end{aligned}
\end{equation}
The covariant derivatives $D\gamma_i$ and $D\beta$ are given by
\begin{subequations}\label{covderivs_gamma_beta}
\begin{align}
\Dc \gamma_i & = \diff \gamma_i - \epsilon_{ij} \tor^j (\gamma_k G^k + \tilde{C}) \ , \\
\Dc \beta & = \diff \beta + \tfrac12 c_{iJ} \tilde{\Tor}_{iI}{}^J C^I \ .
\end{align}
\end{subequations}
The kinetic and topological terms for the vector fields are
\begin{equation}\label{n=4vector}
\begin{aligned}
S_{\mathrm{vector}} = -\tfrac14 \int_{M_{1,3}} & \Big[ \e^{-2\phi-\eta} \tilde{g}_{ij} \Dc G^i \wedge \ast \Dc G^j + \e^{-\eta-\rho} \vert \diff A - a_i \Dc G^i \vert^2 \\
& + \e^{-2\phi+\eta} \tilde{g}^{ij} (\Dc B_i - b_{12} \epsilon_{ik} \Dc G^k) \wedge \ast (\Dc B_j - b_{12} \epsilon_{jl} \Dc G^l) \\
& + \e^{-\eta+\rho} \big\vert \Dc \tilde{C} - \gamma_i \Dc G^i + b_I (\Dc C^I - \tfrac12 b^I \diff A - c_k^I \Dc G^k) \big\vert^2 \\
& + \e^{-\eta} H_{IJ} \big( \Dc C^I - b^I \diff A - c_i^I \Dc G^i \big) \wedge \ast \big( \Dc C^J - b^J \diff A - c_j^J \Dc G^j \big) \\
& + \ b_{12} \eta_{IJ} \Dc C^I \wedge \Dc C^J + 2 b_{12} \diff A \wedge \Dc \tilde{C} \\
& - 2 (\Dc B_i - b_{12} \epsilon_{il} \Dc G^l) \wedge \epsilon^{ij} \big[ (c_{jI} + a_j b_I) \Dc C^I + (\gamma_j - \tfrac12 a_j b_I b^I) \diff A \\
& \hspace{1.2in} {} + a_j \Dc \tilde{C} - (\epsilon_{jk} \beta + a_j \gamma_k + \tfrac12 c_{jI} c_k^I + a_j b_I c_k^I) \Dc G^k \big] \\
& + 2 B_i \wedge \big( \epsilon^{ij} \tilde{\Tor}_{jIJ} C^I \wedge \Dc C^J + \tor^i \tilde{C} \wedge \diff A \big) \Big] \ .
\end{aligned}
\end{equation}
Here, the non-Abelian field-strength for the vector field $\tilde{C}$ is
\begin{equation}\label{DCtilde}
\Dc \tilde{C} = \diff \tilde{C} + \epsilon_{ij} \tor^j G^i \wedge \tilde{C} \ .
\end{equation}
Finally, the  total potential reads
\begin{equation}\label{n=4potential}
\begin{aligned}
S_{\mathrm{potential}}= -\tfrac14 \int_{M_{1,3}} & \Big[ \e^{4\phi+\eta+\rho} \big( \epsilon^{ij} b_I \tilde{\Tor}^I_{iJ} c_j^J \big)^2 + \tfrac52 \e^{2\phi+\eta} \tilde{g}_{ij} \tor^i \tor^j + \tfrac14 \e^{2\phi-\eta+\rho} \tilde{g}_{ij} \tor^i \tor^j H^{IJ} b_I b_J \\
& + \tfrac14 \e^{2\phi+\eta} \tilde{g}^{ij} {[H, \Tor_i]^I}_J {[H, \Tor_j]^J}_I + \e^{2\phi+\eta+\rho} \tilde{g}^{ij} H^{IJ} \Tor^K_{iI} \Tor^L_{jJ} b_K b_L \\
& + \e^{4\phi + \eta} H_{IJ} \big[\epsilon^{ij} \Tor^I_{iK} (c_j^K + a_j b^K) - \tor^i (c_i^I - a_i b^I) \big] \cdot \\
& \hspace{1in} \cdot \big[\epsilon^{kl} \Tor^J_{kL} (c_l^L + a_l b^L) - \tor^k (c_k^J - a_k b^J) \big]
\Big] \ast 1 \ .
\end{aligned}
\end{equation}

\subsection{Determination of the embedding tensor} \label{sec:embedding_tensor}

At this point, we can identify which vector fields in the effective
action~\eqref{Seff} correspond to the electric vector fields $V^{M+}$
in the canonical action~\eqref{N=4general} and which vector fields
should be dualized. Setting the parameters~$\Tor^I_{iJ}$ and~$\tor^i$
to zero in the action~\eqref{Seff}, we can very easily trade vector
fields for their electric-magnetic duals via the usual dualization
procedure. It turns out that exchanging the vector fields~$B_i$ with
their dual fields~$B^{\bar{\imath}}$ suffices to bring the (ungauged)
Lagrangian into the form~\eqref{N=4_canonical_kin}.\footnote{Note that
  turning off the parameters $\Tor^I_{iJ}$ and~$\tor^i$ corresponds to
  compactifications on $K3 \times T^2$. The effective action for this
  case has been determined in \cite{Spanjaard:2008zz,Duff:1995wd,Duff:1995sm}.} 
The computation of the action for the fields~$B^{\bar{\imath}}$ is given in section~\ref{sec:dual-Bi} of the Appendix.

From the action for the dualized fields we can determine the $\so{6,n}$ metric~$\eta_{MN}$ as well as the complex scalar~$\tau$ and the coset matrix~$M_{MN}$ which determine the canonical action~\eqref{N=4_canonical_kin}. If we choose to arrange the electric vectors into the fundamental representation of $\so{6,n}$ as
\begin{equation}\label{electric_vectors}
V^{M+} = (G^i, B^{\bar{\imath}}, A, \tilde C, C^I)
\end{equation}
we find that the $\so{6,n}$ metric $\eta_{MN}$ is given by
\begin{equation}\label{SO6nmetric}
\eta_{MN} = \left(
\begin{array}{ccccc}
0 & \delta_{i\bar{\jmath}} & 0 & 0 &0\\
\delta_{\bar{\imath} j} & 0 & 0 &0 &0\\
0 & 0 & 0 &1 & 0\\
0 & 0 & 1& 0 & 0\\
0 & 0 & 0 & 0 & \eta_{IJ}
\end{array}
\right)\ ,
\end{equation}
and that the scalar factor in the topological vector field couplings is given by
\begin{equation}\label{imtau}
\Re\tau = -\tfrac12 {b_{12}} \ .
\end{equation}
We can find the imaginary part of $\tau$ by checking the kinetic term for $b_{12}$ in the action~\eqref{action_NS}, since according to~\eqref{N=4_canonical_kin} this should contain a factor $(\Im\tau)^{-2}$. In this way, we determine that the complex scalar $\tau$ is given by
\begin{equation}\label{tau}
\tau = \tfrac12 (-b_{12} + \iu \e^{-\eta}) \ .
\end{equation}
For completeness, the matrix $M_{MN}$ is given in Appendix \ref{sec:so6-n-coset}.

We now have enough information to determine the embedding tensor from the covariant derivatives and the non-Abelian field strengths in the action~\eqref{Seff}. We start by determining the components~$\xi_{\alpha M}$ from the covariant derivative of $\tau$. Comparing Eqs.~\eqref{Deta} and~\eqref{Db12} with the general formula~\eqref{Dtau} we conclude that
\begin{equation}\label{embedding_tensor_xi}
\xi_i = - \epsilon_{ij} t^j\ ,
\end{equation}
and $\xi_{\bar{\imath}} = \xi_5 = \xi_6 = \xi_I = 0$. On the other hand, the components $f_{MNP}$ of the embedding tensor are most easily determined from the non-Abelian field strengths of the vector fields~$V^{M+}$. It turns out that setting
\begin{subequations}\label{embedding_tensor_f}
\begin{align}
f_{ij\bar{\imath}} & = - \tfrac12 \epsilon_{ij} \delta_{\bar{\imath}
  k} \tor^k \ ,\\
f_{i56} & = \tfrac12 \epsilon_{ij} \tor^j \ , \\
f_{iIJ} & = - \Tor_{iIJ} \ ,
\end{align}
\end{subequations}
in the general formula~\eqref{fieldstrengths_embedding_tensor} leads
to an agreement with the field-strengths computed in~\eqref{DGi}, \eqref{DCI} and \eqref{DCtilde}. Moreover, it can be checked that~\eqref{embedding_tensor_xi} and \eqref{embedding_tensor_f} satisfy the following quadratic constraints described in Ref.~\cite{Schon:2006kz},
\begin{equation}
\xi^M \xi_M = 0 \ , \qquad \xi^M f_{MNP} = 0 \ , \qquad 3 f_{R[MN} f_{PQ]}{}^R - 2 \xi_{[M} f_{NPQ]} = 0 \ ,
\end{equation}
where square brackets denote antisymmetrization of the corresponding
indices.
That the first two constraints are satisfied follows trivially from the
expressions~\eqref{embedding_tensor_xi} and \eqref{embedding_tensor_f} with
a metric \eqref{SO6nmetric}. The third one follows from 
the commutation relation satisfied by the matrices $T^I_{iJ}$
given in Eq.~\eqref{XXXX}, which as we saw is a consequence of demanding
nilpotency of the exterior differential acting on the two-forms $\omega^I$.

We now have all the information we need in order to write down the action with charged fields in the electric frame. The total field-strength for the electric vector field $B^{\bar{\imath}}$ in the action~\eqref{N=4_canonical_kin} is then
\begin{equation}\label{fieldstrength_B_electric}
F^{\bar{\imath} +} = \diff B^{\bar{\imath}} + \tfrac12 \delta^{i\bar{\imath}} \big[ \epsilon_{ik} \tor^k (\delta_{j\bar{\jmath}} G^j \wedge B^{\bar{\jmath}} - A \wedge \tilde{C}) + \Tor_{iIJ} C^I\wedge C^J - \epsilon_{ij} \tor^j B^{++}\big] \ ,
\end{equation}
while the topological term is given by
\begin{equation}\begin{aligned}\label{topological_action}
S_\mathrm{top} = \tfrac14 \int_{M_{1,3}} & \Big[ B^{++} \wedge \tor^j \Dc B_j - \tor^i \Dc B_i \wedge (\delta_{j\bar{\jmath}} B^{\bar{\jmath}} \wedge \Dc G^j + \tilde{C} \wedge \diff A) \\
& \qquad + 2 \tor^i B_i \wedge (\delta_{j\bar{\jmath}} G^j \wedge B^{\bar{\jmath}} + A \wedge \tilde{C} + \tfrac12 \eta_{IJ} C^I \wedge \Dc C^J) \Big] \ .
\end{aligned}\end{equation}
Using the expressions for $f_{MNP}$, $M_{MN}$ and $\eta_{MN}$, it can
be shown that the potential in \eqref{N=4pot} agrees with the
potential \eqref{n=4potential} obtained from the KK reduction.

Summarizing, we have obtained an action of the form given in \eqref{N=4_canonical_kin}, \eqref{N=4top} and \eqref{N=4pot}. In order to write the action in this form, we had to introduce extra vector fields $B^{\bar{\imath}}$, as well as a tensor field $B^{++}$, which appears in the field strength $F^{+\bar{\imath}}$. To see that this form of the action is equivalent to the action given in equations \eqref{n=4scalar}, \eqref{n=4vector} and \eqref{n=4potential}, one can use the equations of motion for $B^{++}$ to eliminate $B^{++}$ and $B^{\bar{\imath}}$. This reduces the action for the vector fields to the one in \eqref{n=4vector}.

\subsection{Killing vectors and gauge algebra}\label{Killing}

Finally let us determine the gauge group which arises from the
compactifications studied in this paper. It will be  useful
to collectively denote all $(6n+2)$ scalar fields in the effective action by
\begin{equation}
\varphi^\Lambda = (b_{12}, \eta, \phi, \tilde{g}_{ij}, \rho,
\zeta^x_I, a_i, \gamma_i, c_i^I, \beta, b_I)\ , \qquad 
\Lambda = 1, \ldots, 6n+2 \ .
\end{equation}
Then the Killing vectors $k_{M\alpha} = k^\Lambda_{M\alpha}(\varphi) \frac\partial{\partial\varphi^\Lambda}$ can be read off from the covariant derivatives of these fields in Eqs.~\eqref{fieldstrengths_NS}, \eqref{fieldstrengths_RR} and~\eqref{covderivs_gamma_beta} by comparing with the general formula
\begin{equation}
\Dc \varphi^\Lambda = \diff \varphi^\Lambda - k^\Lambda_{M\alpha}(\varphi) V^{M\alpha} \ .
\end{equation}
Doing this, we obtain the following expressions for the Killing vectors,
\begin{equation}
\begin{aligned}
k_{i+} = {} & \epsilon_{ij} \tor^j \Big( b_{12} \frac\partial{\partial b_{12}} - \frac\partial{\partial \eta} + \frac\partial{\partial \rho} \Big) - \Tor^J_{iI} \zeta^x_J \frac\partial{\partial \zeta^x_I} + \tor^j (\epsilon_{ik} \tilde{g}_{jl} + \epsilon_{il} \tilde{g}_{jk} - \epsilon_{ij} \tilde{g}_{kl}) \frac\partial{\partial \tilde{g}_{kl}} \\
& \qquad + \epsilon_{ij} \tor^k a_k \frac\partial{\partial a_j}+ \epsilon_{jk} \tor^k \gamma_i \frac\partial{\partial \gamma_j} + \big(\epsilon_{ij} \tor^k \delta_I^J - \delta_j^k \tilde{\Tor}_{iI}{}^J \big) c_k^I \frac\partial{\partial c_j^J} - \tilde{\Tor}^J_{iI} b_J \frac\partial{\partial b_I} \ , \\
k_{6+} = {} & \epsilon_{ij} \tor^j \frac\partial{\partial \gamma_i} \ , \qquad\quad
k_{I+} = {} \tilde{\Tor}_{iI}{}^J \Big(\frac\partial{\partial c_i^J} - \tfrac12 \epsilon^{ij} c_{jJ} \frac\partial{\partial \beta} \Big) \ , \qquad
k_{i-} = {} \epsilon_{ij} \tor^j \frac\partial{\partial b_{12}} \ , \\
k_{\bar{\imath}\pm} = {} & k_{5\pm} = k_{6-} = k_{I-} = 0 \ .
\end{aligned}
\end{equation}
Now we can compute the Lie brackets for this set of vectors to obtain
\begin{equation}\label{Lie_brackets}
\begin{aligned}
{}[k_{i+}, k_{j+}] & = -\epsilon_{ij} \tor^k k_{k+} \ , \qquad [k_{i+}, k_{6+}] = -\epsilon_{ij} \tor^j k_{6+} \ , \\
[k_{i+}, k_{I+}] & = -\tilde{\Tor}_{iI}{}^J k_{J+} \ , \qquad [k_{i+}, k_{j-}] = \epsilon_{jk} \tor^k k_{i-} \ ,
\end{aligned}
\end{equation}
with the all other brackets vanishing. 
Inspecting \eqref{differential_algebra} we
see that by choosing appropriate linear combinations of $v^1$ and $v^2$
we can set $\tor^1=0$ without loss of generality and then rename  $\tor^2\equiv\tor$.  If we do this, $k_{2-}$ is zero, and the non-vanishing Lie brackets~\eqref{Lie_brackets} read
\begin{equation}
\begin{aligned}
{}[k_{1+}, k_{2+}] & = \tor k_{2+} \ , \qquad [k_{1+}, k_{1-}] = -\tor k_{1-} \ , \\
[k_{1+}, k_{6+}] & = \tor k_{6+} \ , \qquad [k_{1+}, k_{I+}] = \tfrac12 \tor k_{I+} + \Tor^J_{1I} k_{J+} \ ,  \\
[k_{2+}, k_{I+}] & = \Tor^J_{2I} k_{J+} \ .
\end{aligned}
\end{equation}
This corresponds to the solvable algebra $(\mathbb{R}_{k_{6+}} \times \mathbb{R}_{k_{1-}} \times (\mathbb{R}^n_{k_{I+}} \rtimes \mathbb{R}_{k_{2+}})) \rtimes \mathbb{R}_{k_{1+}}$, where in the first semi-direct product, $\mathbb{R}_{k_{2+}}$ acts on $\mathbb{R}^n_{k_{I+}}$ by means of the matrix $\Tor^J_{2I}$, while in the second, $\mathbb{R}_{k_{1+}}$ acts on $\mathbb{R}_{k_{6+}} \times \mathbb{R}_{k_{1-}} \times (\mathbb{R}^n_{k_{I+}} \rtimes \mathbb{R}_{k_{2+}})$ through the matrix
\begin{equation}
\mathrm{diag} (\tor, -\tor, \tfrac12 \tor \delta^J_I + \Tor^J_{1I}, \tor) \ .
\end{equation}

That the algebra \eqref{Lie_brackets} is indeed consistent with gauged $\mathcal{N} = 4$ supergravity we see by defining the following matrices~\cite{Schon:2006kz}
\begin{equation}\label{X_matrices}
X_{M+} = \begin{pmatrix}
X_{M+N+}{}^{P+} & 0 \\
0 & X_{M+N-}{}^{P-}
\end{pmatrix} \ , \qquad
X_{M-} = \begin{pmatrix}
0 & X_{M-N+}{}^{P-} \\
0 & 0
\end{pmatrix} \ ,
\end{equation}
with non-vanishing entries given in terms of the embedding tensors by
\begin{equation}
\begin{aligned}
X_{M+N+}{}^{P+} & = -f_{MN}{}^P - \tfrac12 (\delta^P_M \xi_N - \delta^P_N \xi_N - \eta_{MN} \xi^P) \ , \\
X_{M+N-}{}^{P-} & = -f_{MN}{}^P - \tfrac12 (\delta^P_M \xi_N + \delta^P_N \xi_N - \eta_{MN} \xi^P) \ , \\
X_{M-N+}{}^{P-} & = - \delta^P_N \xi^M \ .
\end{aligned}
\end{equation}
As discussed in Ref.~\cite{Schon:2006kz}, the non-Abelian gauge algebra of the $\mathcal{N} = 4$ supergravity should be reproduced by the commutators
\begin{equation}\label{comm_X}
\begin{aligned}
{} [X_{M+}, X_{N+}] & = X_{M+N+}{}^{P+} X_{P+} \ , \\
[X_{M+}, X_{N-}] & = X_{M+N-}{}^{P-} X_{P-} = -X_{N-M+}{}^{P-} X_{P-} \ , \\
[X_{M-}, X_{N-}] & = 0 \ , \\
\end{aligned}
\end{equation}
And indeed, by using the expressions \eqref{embedding_tensor_xi} and \eqref{embedding_tensor_f} for the embedding tensor in the formulas \eqref{X_matrices} to \eqref{comm_X}, the algebra \eqref{Lie_brackets} is recovered.

\section{Conclusions}\label{section:Conclude}

In this paper we considered type IIA supergravity compactified on
a specific class of six-dimensional manifolds which have $\su2$
structure. Such manifolds admit a pair of globally defined spinors and
they can be further characterized by their non-trivial intrinsic
torsion. Among the $\su2$-structure manifolds one also finds the
Calabi-Yau manifold $K3\times T^2$ for which  the intrinsic torsion
vanishes. Furthermore, the entire class of six-dimensional
$\su2$-structure manifolds  necessarily has an almost product
structure of a four-dimensional component times a two-dimensional
component which also generalizes the Calabi-Yau case. However, in
order to simplify the analysis in this paper, we confined our
attention to torsion classes which lead to an integrable almost 
product structure.

For this class of compactifications (with the additional requirement
of the absence of massive gravitino multiplets) we determined the
resulting four-dimensional $\mathcal{N}=4$ low-energy effective action
by performing a Kaluza-Klein reduction. By appropriate dualizations of
one- and two-forms it was possible to go from the `natural' field
basis of the KK reduction to a supergravity field basis where the
consistency with  the `standard' $\mathcal{N}=4$ form as given in
\cite{Schon:2006kz} could be established. In that process, we
determined the components of the embedding tensor or in other words
the couplings of the $\mathcal{N}=4$ action in terms of the intrinsic
torsion. The resulting gauge group is solvable, as usually is the case 
for these compactifications.

\section*{Acknowledgments}

B.S.~would like to acknowledge useful discussions with Stefan Groot-Nibbelink, Olaf Hohm, Andrei Micu, Ron Reid-Edwards, Henning Samtleben and Martin Weidner.

This work was partly supported by the German Science Foundation (DFG)
under the Collaborative Research Center (SFB) 676 ``Particles, Strings
and the Early Universe''.
The work of H.T.\ is supported by the DSM CEA/Saclay, the ANR grant 08-JCJC-0001-0 and the ERC Starting
Independent Researcher Grant 240210 - String-QCD-BH. 

\vfill

\newpage

\appendix

\noindent
{\bf\Large Appendix}
\section{Dualizations} \label{sec:dualizations-appendix}

In this appendix we give some of the calculational steps involved in the field dualizations from Sections~\ref{sec:dualizations} and~\ref{sec:embedding_tensor}. The purpose of a field dualization is to obtain an equivalent theory, where a (massless) $p$-form field is exchanged for a $(2-p)$-form field.

\subsection{Dualization of two-forms} \label{sec:dual-two-forms}

The Lagrangian obtained from the compactification still contains the two-forms $B$ and $C_i$, which we can exchange for vector and scalar fields by performing the appropriate dualizations \cite{Louis:2002ny}.

We start by dualizing the two-form fields $C_i$ and the one-form
$C_{12}$. Collecting all the relevant terms obtained from the action
\eqref{action_RR_kin} and 
\eqref{action_RR_CS} for the RR fields, we have
\begin{equation}\begin{aligned}\label{action_CiC12}
S_{C_i, C_{12}} = -\tfrac14 \int_{M_{1,3}} & \Big[ \e^{-2\phi-\rho} \tilde{g}^{ij}(H_i + J_i) \wedge \ast (H_i + J_i) + \e^{\eta-\rho} \vert F_{12} + J_{12} \vert^2 \\
& - 2 H_i \wedge \epsilon^{ij} b_I \Dc c_j^I + 2 F_{12} \wedge K \Big] \ ,
\end{aligned}\end{equation}
where, for simplicity, we have introduced the following abbreviations,
\begin{equation}
\begin{aligned}
H_i & \equiv \Dc C_i =  \diff C_i + \epsilon_{ij} \tor^k G^j \wedge C_k - \epsilon_{ij} C_{12} \wedge \Dc G^j \ , \\
F_{12} & \equiv \Dc C_{12} = \diff C_{12} + \tor^i C_i + \epsilon_{ij} \tor^j G^i \wedge C_{12} \ , \\
J_i & = a_i \Dc B - \diff A \wedge B_i \ , \\
J_{12} & = - b_{12} \diff A - a_i ( \epsilon^{ij} \Dc B_j - b_{12} \Dc G^i) \ , \\
K & = b_I (\Dc C^I - \tfrac12 b^I \diff A - c_k^I \Dc G^k) \ .
\end{aligned}
\end{equation}
The fact that the bare $p$-form potential $C_i$ also appears in the field strength $F_{12}$ makes it impossible to replace $C_i$ by dual scalar fields $\gamma_i$ without also replacing $F_{12}$ by the field strength of a dual vector field $\tilde{C}$.  We can do this by constructing an action which is equivalent to~\eqref{action_CiC12}, where the field strengths $H_i$ and $F_{12}$ are treated as independent fields. The equivalence to the original Lagrangian is guaranteed by introducing Lagrange multipliers $\gamma_i$ and $\tilde{C}$ which enforce the correct Bianchi identities for $H_i$ and $F_{12}$, namely
\begin{equation}\label{bianchi_identities}
\begin{aligned}
\diff H_i & = - \epsilon_{ij} \tor^k G^j \wedge H_k + \epsilon_{ik} \Dc G^k \wedge F_{12} \ ,\\
\diff F_{12} & = \tor^i H_i + \epsilon_{ij} \tor^j G^i \wedge F_{12} \ .
\end{aligned}
\end{equation}
The modified action thus becomes
\begin{equation}\label{action_CiC12_modified}
\begin{aligned}
S^\prime_{C_i,C_{12}} & = S_{C_i,C_{12}} - \tfrac12 \int_{M_{1,3}} \Big[ \gamma_i \epsilon^{ij} (\diff H_j + \epsilon_{jk} \tor^l G^k \wedge H_l - \epsilon_{jk} \Dc G^k \wedge F_{12}) \\
& \hspace{2in} {} + \tilde{C} \wedge (\diff F_{12} - \tor^i H_i - \epsilon_{ij} \tor^j G^i \wedge F_{12}) \Big] \\
& = S_{C_i,C_{12}} + \tfrac12 \int_{M_{1,3}} \Big[ H_i \wedge \epsilon^{ij} (\diff \gamma_j + \epsilon_{jk} \tor^k \gamma_l G^l + \epsilon_{jk} \tor^k \tilde{C}) \\
& \hspace{2in} {} - F_{12} \wedge (\diff\tilde{C} + \epsilon_{ij} \tor^j G^i \wedge \tilde{C} - \gamma_i \Dc G^i) \Big] \ . \\
\end{aligned}
\end{equation}
Integrating out the fields $H_i$ and $F_{12}$ by using their equations of motion leads to the following action for the dual fields $\gamma_i$ and $\tilde{C}$,
\begin{equation}\begin{aligned}\label{action_CiC12_dual}
S_{\tilde{C},\gamma_i} = -\tfrac14 \int_{M_{1,3}} & \Big[ \e^{2\phi+\rho} \tilde{g}^{ij} (\Dc \gamma_i + b_I \Dc c_i^I) \wedge \ast (\Dc \gamma_j + b^J \Dc c_j^J) \\
& + e^{-\eta+\rho} \vert \Dc \tilde{C} - \gamma_i \Dc G^i + K \vert^2 \\
&+ 2 \epsilon^{ij} (D\gamma_i + b_I D c_i^I) \wedge J_j - 2 (\Dc \tilde{C} - \gamma_i \Dc G^i + K) \wedge J_{12} \Big] \ ,
\end{aligned}\end{equation}
where we have defined the covariant derivatives $\Dc \gamma_i$ and the non-Abelian field-strength $\Dc \tilde{C}$ as
\begin{align}
\Dc \gamma_i & = \diff \gamma_i - \epsilon_{ij} \tor^j (\gamma_k G^k + \tilde{C}) \ , \\
\Dc \tilde{C} & = \diff \tilde{C} + \epsilon_{ij} \tor^j G^i \wedge \tilde{C} \ .
\end{align}

The dualization of the two-form $B$ is much simpler, due to the simpler nature of its couplings. After the dualization of the two-forms $C_i$, the action for $B$, written in terms of its field strength $H \equiv \Dc B = \diff B + B_i \wedge \Dc G^i$ and introducing a Lagrange multiplier $\beta$ to enforce $\diff^2 B = \diff (H - B_i \wedge \Dc G^i) = 0$, is given by
\begin{equation}\label{action_B}
S_B = -\tfrac14 \int_{M_{1,3}} \big[ \e^{-4\phi} \vert H \vert^2 + 2 H \wedge W + 2 \beta \wedge \diff (H - B_i \wedge \Dc G^i) \big] \ ,
\end{equation}
with the shorthand
\begin{equation}
W = \epsilon^{ij} (a_i D\gamma_j + a_i b_I \Dc c_j^I - \tfrac12 c_{iI} \Dc c_j^I + \tfrac12 c_{iJ} \tilde{\Tor}_{iI}{}^J C^I) \ .
\end{equation}
Eliminating $H$ by using its equations of motion, we obtain the action for the dual scalar field $\beta$,
\begin{multline}\label{action_dual_B}
S_\beta = -\tfrac14 \int_{M_{1,3}} \Big[ \e^{4\phi} \big\vert \Dc \beta - \epsilon^{ij} (a_i \Dc \gamma_j + a_i b_I \Dc c_j^I - \tfrac12 c_{iI} \Dc c_j^I) \big\vert^2 - 2 \beta \wedge \Dc B_i \wedge \Dc G^i \Big] \ ,
\end{multline}
where the covariant derivative of $\beta$ is
\begin{equation}
\Dc \beta = \diff \beta + \tfrac12 c_{iJ} \tilde{\Tor}_{iI}{}^J C^I \ .
\end{equation}

\subsection{Finding the correct electric-magnetic duality frame}\label{sec:dual-Bi}

In order to read off the gauge couplings $M_{MN}$ and $\eta_{MN}$, we can consider the action with all charges $T^I_{iJ}$ and $t^i$ set to zero, and bring this action into the correct electric-magnetic duality frame. When no fields are charged with respect to the vector fields, the dualizations are of course simpler, and we find that replacing the vector fields $B_i$ by their duals $B^{\bar{\imath}}$ brings the couplings into their canonical form.

Setting charges to zero, the terms in the action containing the fields $B_i$ are
\begin{equation}\begin{aligned}\label{action_Bi}
S_{B_i} = - \tfrac14 \int_{M_{1,3}} & \Big[ \e^{-2\phi+\eta} \tilde{g}^{ij} (F_i - b_{12} \epsilon_{ik} \diff G^k) \wedge \ast (F_j - b_{12} \epsilon_{jl} \diff G^l) \\
& - 2 (F_i - b_{12} \epsilon_{ik} \diff G^k) \wedge \epsilon^{ij} L_j \Big] \ ,
\end{aligned}\end{equation}
where $F_i = \diff B_i$ and we have introduced the shorthand notation
\begin{equation}\label{abbrevLi}
\begin{aligned}
L_i  =  & (c_{iI} + a_i b_I) \Dc C^I + (\gamma_i - \tfrac12 a_i b_I b^I) \diff A + a_i \Dc \tilde{C} \\
&  - (\epsilon_{ij} \beta + a_i \gamma_j + \tfrac12 c_{iI} c_j^I +  a_i b_I c_j^I) \Dc G^j \ .
\end{aligned}
\end{equation}
We now introduce the dual fields $B^{\bar{\imath}}$ by adding the following term to the action \eqref{action_Bi},
\begin{equation}\label{lagrange_dual_Bi}
\delta S = - \tfrac12 \int_{M_{1,3}} \delta_{\bar{\imath}i} B^{\bar{\imath}} \wedge \epsilon^{ij} \diff F_j \ .
\end{equation}
Eliminating the two-forms $F_i$ using its equations of motion, we arrive at the dual action
\begin{multline}\label{action_dual_Bi}
S_{B^{\bar{\imath}}} = -\tfrac14 \int_{M_{1,3}} \Big[ \e^{2\phi-\eta} \tilde{g}^{ij} (\delta_{i\bar{\imath}} \diff B^{\bar{\imath}} + L_i) \wedge \ast (\delta_{j\bar{\jmath}} \diff B^{\bar{\jmath}} + L_j) + 2 b_{12} \delta_{\bar{\imath}i} \diff  B^{\bar{\imath}}\wedge \diff G^i \Big] \ .
\end{multline}
Substituting these results into the complete vector action~\eqref{n=4vector}, we can see that the gauge kinetic couplings are now indeed of the canonical form presented in equation \eqref{N=4_canonical_kin}. This allows us to read off the matrices $M_{MN}$ and $\eta_{MN}$.

\section{$\mathrm{SO}(6,n)$ coset matrix}
\label{sec:so6-n-coset}

The entries of the matrix $M_{MN}$ (with indices $M,N$ taking the $6+n$ values $i$, $\bar{\imath}$, $5$, $6$, $I$) can be easily extracted from the kinetic terms for the vectors in Eqs.~\eqref{n=4vector} and~\eqref{action_dual_Bi}, by comparison with the general form of this term for $\mathcal{N} = 4$ supergravity in Eq.~\eqref{N=4_canonical_kin}. The result is
\begin{align}
M_{ij} & = \e^{-2\phi} \tilde{g}_{ij} + \e^{-\rho} a_i a_j + \e^\rho (\gamma_i + b_I c_i^I) (\gamma_j + b_J c_j^J) + H_{IJ} c_i^I c_j^J \nonumber \\
& \quad {} + \e^{2\phi} \tilde{g}^{kl} (\epsilon_{ki} \beta + a_k \gamma_i + \tfrac12 c_{kI} c_i^I + a_k b_I c_i^I) (\epsilon_{lj} \beta + a_l \gamma_j + \tfrac12 c_{lI} c_j^I +a_l b_I c_j^I) \ , \\
M_{i\bar{\jmath}} & = \e^{2\phi} \tilde{g}^{jk} \delta^j_{\bar{\jmath}} (\epsilon_{ki} \beta + a_k \gamma_i + \tfrac12 c_{kI} c_i^I + a_k b_I c_i^I) \ , \\
M_{i5} & = - \e^{-\rho} a_i + \e^{\rho} b_I b^I (\gamma_i + b_J c_i^J) + H_{IJ} b^I c_i^J \\
& \qquad \qquad {} + \e^{2\phi} \tilde{g}^{jk} (\gamma_j - \tfrac12 a_j b_I b^I) (\epsilon_{ki} \beta + a_k \gamma_i + \tfrac12 c_{kI} c_i^I +  a_k b_I c_i^I) \ , \\
M_{i6} & = - \e^\rho (\gamma_i + b_I c_i^I) - \e^{2\phi} \tilde{g}^{jk} a_j (\epsilon_{ki} \beta + a_k \gamma_i + \tfrac12 c_{kI} c_i^I + a_k b_I c_i^I) \ , \\
M_{iI} & = - H_{IJ} c_i^J - \e^\rho b_I (\gamma_i + b_J c_i^J) \\
& \qquad \qquad  {} + \e^{2\phi} \tilde{g}^{jk} (c_{jI} + a_j b_I) (\epsilon_{ki} \beta + a_k \gamma_i + \tfrac12 c_{kI} c_i^I + a_k b_I c_i^I) \ , \\
M_{\bar{\imath}\bar{\jmath}} & = \e^{2\phi} \tilde{g}^{ij} \delta_{i\bar{\imath}} \delta_{j\bar{\jmath}} \ , \\
M_{\bar{\imath}5} & = \e^{2\phi} \tilde{g}^{ij} \delta_{i\bar{\imath}} (\gamma_j - \tfrac12 a_j b_I b^I) \ , \\
M_{\bar{\imath}6} & = - \e^{2\phi} \tilde{g}^{ij} \delta_{i\bar{\imath}} a_j \ , \\
M_{\bar{\imath}I} & = \e^{2\phi} \tilde{g}^{ij} \delta_{i\bar{\imath}} (c_{jI} + a_j b_I) \ , \\
M_{55} & = \e^{-\rho} + \tfrac14 \e^\rho (b_I b^I)^2 + \e^{2\phi} \tilde{g}^{ij} (\gamma_i - \tfrac12 a_i b_I b^I) (\gamma_j - \tfrac12 a_j b_J b^J) + H_{IJ} b^I b^J \ , \\
M_{56} & = - \tfrac12 \e^\rho b_I b^I - \e^{2\phi} \tilde{g}^{ij} a_i (\gamma_j - \tfrac12 a_j b_I b^I) \ , \\
M_{5I} & = - \tfrac12 \e^\rho b_I b^I b_J - H_{IJ} b^J + \e^{2\phi} \tilde{g}_{ij} (c_{iI} + a_i b_I) (\gamma_j - \tfrac12 a_j b_J b^J) \ , \\
M_{66} & = \e^\rho + \e^{2\phi} \tilde{g}^{ij} a_i a_j \ , \\
M_{6I} & = \e^\rho b_I - \e^{2\phi} \tilde{g}^{ij} a_i (c_{jI} + a_j b_I) \ , \\
M_{IJ} & = H_{IJ} + \e^\rho b_I b_J + \e^{2\phi} \tilde{g}^{ij} (c_{iI} + a_i b_I) (c_{jJ} + a_j b_J) \ ,
\end{align}
with the other entries determined by symmetry. It can be checked that this matrix indeed satisfies $M_{MN} \eta^{NP} M_{PQ} = \eta_{MQ}$, with the $\so{6,n}$ metric given in Eq.~\eqref{SO6nmetric}.


\end{document}